\documentclass[runningheads]{llncs}
\usepackage{graphicx}

\begin{document}
\title{The Intelligence College in Europe (ICE)}
\subtitle{An Effort to Create a European Intelligence Community}
\titlerunning{The Intelligence College in Europe}
\author{Uwe M. Borghoff \inst{1}\thanks{Corresponding author: uwe.borghoff@unibw.de}
\and Lars Berger \inst{2} 
\and François Fischer \inst{3}
}
\authorrunning{Uwe M. Borghoff, Lars Berger \& François Fischer}

\institute{Center for Intelligence and Security Studies (CISS) \& Dept.\ of Computer Science \\ University of the Bundeswehr Munich, Neubiberg, Germany
\and
Department of Intelligence \\ Federal University of Administrative Sciences, Berlin, Germany \\
\and
ICE’s Permanent Secretariat, Paris, France \\
\url{https://www.intelligence-college-europe.org} 
}
\maketitle              

\begin{abstract}
In fulfilling the European security commitment, the actors of the so-called ``Intelligence Community'' play a central role. They provide political and military decision-makers with important analyses and information. The Intelligence College in Europe (ICE) is the first entity to offer professional intelligence training as well as postgraduate level academic education in intelligence and security studies at a pan-European level. In developing its postgraduate provision, ICE has benefited from the experience of the German Master of Intelligence and Security Studies (MISS), which is a joint effort of the University of the Bundeswehr Munich and the Department of Intelligence at the Federal University of Administrative Sciences in Berlin. As a main contribution of this paper, the module Counterterrorism (adapted from the MISS) is examined in more detail as a case study of how postgraduate modules can be modified to speak to a pan-European audience of intelligence professionals.

\end{abstract}

\section*{Introduction to ICE -- The big picture}
The Intelligence College in Europe (ICE) started with a vision of a ``European Intelligence Academy'' expressed by French President Macron in his speech at Sorbonne in September 2017, gathering for the first time European national intelligence communities to work together in a non-operational way. So far, the European intelligence community was fragmented or non-existent at all, refer, for instance, to \cite{prunckun2016handbook}.
Inaugurated in 2019, the Intelligence College in Europe is an intergovernmental entity, distinct from the European Union, aiming to enhance European security and to build a shared strategic culture for intelligence in Europe, bringing together practitioners and academics. The ICE also enables the European intelligence community to enter into a dialogue with European decision-makers, intelligence consumers, and civil society. 
In 2024, ICE will celebrate its 5th anniversary. Already now, on the eve of this event, we can point to significant progress in three domains:

\begin{enumerate}
\item
30 countries and their national intelligence communities are part of ICE: 27 member countries (of which 22 are EU member states) and three partner countries, gathering some 70 intelligence and security services.
\item
ICE’s academic network now consists of 33 academic institutions, universities and think tanks in 18 different member countries.
\item
In the academic year 2023/24 , ICE will offer more than 30 events, training sessions and outreach activities within a clearly structured, high-quality and popular academic programme.
\end{enumerate}

This level of success was not evident from the start: ICE, born out of the idea to foster a much needed common strategic culture, had to try -- contrary, for instance, to the US intelligence community, where convergence ``only'' has to be achieved among intelligence services from the same country -- to develop some common intelligence “strategic culture” with 89 services from 30 countries.

\section*{ICE’s academic points of efforts}
ICE’s activities, of various academic density, but always encompassing a strong academic component are structured around three pillars:

\begin{itemize}
\item
Firstly, {\em thematic seminars}, where intelligence services cadres and experts from the public sphere meet for two to three days. These type of thematic seminars have established themselves as an ICE ``flagship''. Organizing countries, sometimes teaming their efforts, aim to bring together academic research and practitioners’ expertise to demonstrate their countries’ capacities or to put a spotlight on a topic of particular importance for them, in a sound ``influence'' approach. 

It is noteworthy that more and more frequently different countries are bringing together their best experts to offer in-depth ``top-level'' seminars on topics of great European interest.
\item
Secondly, we conduct {\em outreach activities to raise awareness} of the issue. We pursue three main channels: (i) the academic world (with a particular focus on students at Master's or PhD stage as highly qualified participants), (ii) the EU institutions and (iii) publication on our website or in professional journals.
\item
Finally, within our {\em academic programme}, where we offer training and academic teaching delivered by academics and practitioners for intelligence practitioners from our member countries. This programme consists itself of two different formats: (i) the {\em executive education programme} for one set of high potential cadres with five sessions of one week each held in different member countries. (ii) the {\em postgraduate programme}, where participants can attend modules related to a specific field of intelligence studies such as on counter\-terrorism as explained below. In this case, the target audience are analysts at every stage of their career and intelligence professionals who interact and cooperate in clubs or in multilateral institutions (EU, NATO, etc.).
\end{itemize}

What do we want to achieve through education? Firstly, to foster a ``reflex'' of cooperation, a knowledge of the others and an understanding of differences in order to facilitate multinational cooperation on topics of common interests. 

\section*{The three main challenges ICE had to tackle}
The challenges faced while putting together ICE and make it run are the challenges that every European intelligence education initiative faces.

\begin{itemize}
\item
The first challenge we faced was reluctance. Putting together two or more intelligence services was already tricky, but adding academia to the table was the real challenge.
When we speak about European intelligence education we have to speak about a major constrain: contrary to what is going on in a heavily institutionalized military organization such as NATO, we don’t have a ``big brother'' who could, in case of disunion or lack of prioritization, bring us back ``on track''.

So, we truly need a real consensus and, in order to achieve it, a proven, permanent, genuine respect for our diversity and for our various national priorities. Think about issues such as immigration or Africa, they have not the same degree of priority as Ukraine or Russia for all the different European countries. Building consensus is paramount while building a European intelligence offer. This ``diplomatic aspect'' was and is still at the heart of our mission. 

Having doubled the number of courses on offer in the last two years, we are now entering a new phase, an ``appropriation phase'', which is also characterized by the fact that some courses are offered regularly, that we continue to strive for the inherent systematic quality assurance and that new countries are interested in joining the original 30 founding members.
\item
The second big challenge, not linked by one way or another to trust, was the confrontation of will and opportunity. 
It is one of the least ``condemnable'' challenge as there will always be something more important or urgent on the agenda. There are external factors that may change one enterprise’s priorities such as the Russian invasion of Ukraine. Every intelligence organization’s priority shifted massively towards this threat leading to the (re-) allocation of funds and human resources.

On the positive side, this has also improved our activities. The general threat to our common security has brought the intelligence world to the realization that working together is better than going it alone. ICE has more and more spontaneous proposals and is now seeing a strong willingness to develop what were previously seen as sensitive areas of work, such as academic cooperation within our ICE network or the dedicated EU support area.
And, of course, we have to take as a structuring point, that there will never be a pan-European harmonization of the way the various national intelligence communities are established and functioning. So, diversity is a ``given'' in every European intelligence education effort. One unsaid prerequisite in this domain is to accept to make full use of this complex and scattered scene. 
\item
The third big challenge and a major next step is the consolidation of the growth of our academic activities. Remember, the ICE´s main aim is to enhance European security. The ICE thus organized seminars, academic activities and outreach events on many issues of major importance for the European security, such as new security in Europe, strategic communication, innovation and intelligence, strategic analysis in Europe, space in Europe, European security challenges, seminars on strategic competitors such as Russia or China, hybrid threats, countering manipulation of information, cyber defence, OSINT, anticipation by intelligence, disinformation, or military intelligence. Other activities focused on important subjects related to the European security such as counterterrorism or radicalization in our societies.
\end{itemize}

The ICE is steadily enhancing its outreach activities toward the European Union. For the last two years, the ICE has sought to substantially deepen the cooperation with European partners, including the Commission, the Collège d’Europe\footnote{https://www.coleurope.eu/fr}, SIAC\footnote{See also the explanations in \cite{nomikos2014european}}, EU SATCEN\footnote{https://www.satcen.europa.eu/} or the four EU Security branches. 
By this, the ICE trains intelligence customers at the European level, making it an actor in the European security architecture. In 2023, the ICE had already partaken in four events in Brussels, Bruges (Collège d’Europe) or with a strong EU presence. Furthermore, selected EU officials have been invited as attendees to outreach and thematic seminars. In the near future, the ICE intends to create with some help from the European Security \& Defence College (ESDC)\footnote{https://esdc.europa.eu/} a dedicated ``EU line of effort''.

In the following section, we take a look at the academic intelligence education in Germany. The Master of Intelligence and Security Studies (MISS) and the export of two German modules from the MISS to the academic ICE programme are examined in some detail. The counter-terrorism module serves as a case study and is presented in detail.

\section*{Germany’s Master of Intelligence and Security Studies}
Academic intelligence education was quite neglected until 10 years ago and only taught in a few institutions \cite{glees2018intelligence}. In Germany, after a long period of preparation, numerous internal consultations within the participating federal departments and agencies, and in-depth research \cite{Scheffler2016}, the German Federal University of Administrative Sciences and the University of the Bundeswehr Munich have concluded a cooperation agreement, thus sealing their collaboration on the Master's degree programme ``Intelligence and Security Studies'' -- MISS\footnote{This MISS should not be confused with the Master of Arts in {\bf International} Security Studies (also abbreviated MISS), another unique postgraduate programme for security professionals offered jointly by the University of the Bundeswehr Munich and the George C. Marshall European Center for Security Studies (GCMC), see also https://www.marshallcenter.org/en/academics/master-arts-international-security-studies-miss.}.
It is offered as a consecutive attendance course of study with a duration of two years. The academic degrees awarded are -- depending on the field of study -- a ``Master of Arts'' (M.A.) or a ``Master of Science'' (M.Sc.) in the field of ``Intelligence and Security Studies'' jointly by both universities\footnote{https://www.unibw.de/ciss/miss}.

\begin{figure}[t]
\centering\includegraphics[width=\textwidth]{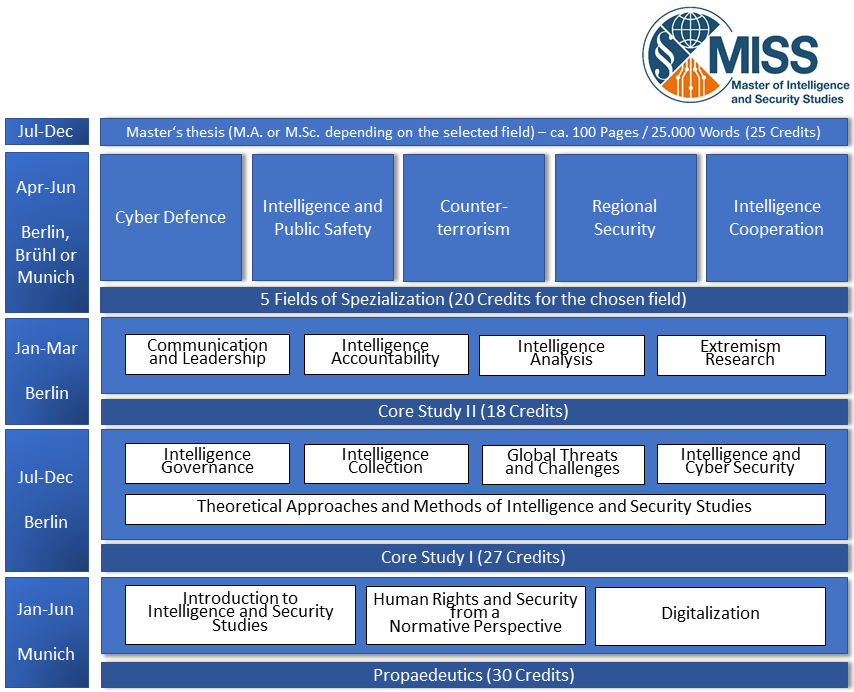}
\caption{The MISS curriculum in brief (the timeline is from bottom to top, i.e.\ from the preparatory course (Propaedeutics) to the Master's thesis).}\label{misscurriculum}
\end{figure}

The study programme, designed and implemented by Borghoff and Dietrich \cite{BorghoffD17}, follows a transdisciplinary approach; in terms of content, the MISS is oriented towards the professional needs of practice. In the course of study, security-relevant facts, problems and developments are taken up from a wide variety of scientific perspectives, e.g.\ law, psychology, political science, computer science, history and sociology. At the same time, as depicted in Figure~\ref{misscurriculum}, economics, media and cultural studies are also part of the programme. In addition to intelligence and military practice, students are taught both technical and methodological, social and personal skills.

Currently, the MISS as a whole is for German eyes only. It is exclusively reserved for current and future employees of the federal intelligence services (BND, BfV and MAD), the federal states (LfV) as well as soldiers and civilian members of the Bundeswehr (especially from the MilNW sector). In addition, the course is open to members of the ministerial administration with links to security policy (especially the Federal Chancellery, the Federal Ministry of the Interior, the Federal Ministry of Defense and the Federal Foreign Office) and to members of the parliamentary administration working in the parliamentary control of the intelligence services.

Germany has supported the ICE and in particular the academic ICE programme from the very beginning. Two modules from the MISS were quickly identified as relevant and suitable for export to the ICE academic programme, or more specifically, to the ICE postgraduate programme. 

\section*{The postgraduate programme as part of the ICE academic programme}
The postgraduate programme is one of two central formats offered within the academic programme of ICE. It complements the executive education programme with a focus on exposing European intelligence professionals to postgraduate-level academic teaching that is informed by cutting-edge research. In providing time and space to critically reflect upon and discuss topics of relevance to Europe’s internal and external security, the postgraduate programme offers a range of benefits. First, the free exchange of ideas deepens the mutual understanding of threat perceptions of individual countries and agencies and facilitates the development of a common European strategic vision among Europe’s future intelligence leaders. Second, the open atmosphere of an academic classroom offers intelligence professionals the opportunity to engage in the kind of ``outside-the-box''-thinking that is crucial in robust intelligence analysis. Third, through the delivery of teaching at academic institutions, the postgraduate programme fosters mutually beneficial links between the academic and intelligence worlds thereby strengthening the developing field of intelligence studies. The emerging network of academics and academic institutions contributing to the postgraduate programme provides institutional memory and thus helps facilitate ICE’s institutionalization as the go-to-place for excellence in research and education in intelligence and security studies at a pan-European level.

The module on {\em Counterterrorism} (adapted from the version taught within MISS) was the first module to be offered under ICE’s postgraduate programme. It has since been followed by the successful launch of the second German module {\em Cyberintelligence and its Implications for Intelligence, Analysis and Decision Making} led by Gerhard Conrad and Stefan Pickl. In this second module, participants gain a deeper understanding of the interrelation and interdependence of the cyber-dimension with political, military and security related decision making processes. How can cyber-based applications contribute to timely and comprehensive situational awareness which is a prerequisite for competent decision making? Additional emphasis is laid on the vital importance of procedures such as anonymizing and disguising communication that make it difficult or impossible to gather intelligence. Basics of applied cryptography, data mining, system dynamics, and interdiction games are introduced that participants can then apply in their respective areas of responsibility. The aim is thus to develop an overall innovative system-related understanding, to identify the essential framework conditions and influencing factors for intelligence and security services as well as their decision-makers in the global cyber world and to promote the ability of modelling and simulating their mode of operation. This interdisciplinary approach combines IT, operations research and policies and transcends conventional intelligence or cyber courses by promoting the concept of ``Support to Decision Making in the Cyber World''.

At the beginning of the academic year 2023/24, the postgraduate programme took further shape with Sebastiaan Rietjens delivering a very-well received module with participants from 15 countries on {\em Intelligence and the Military} at the Netherlands Defence Academy. Taking the Russian invasion of Ukraine in 2022 and the NATO withdrawal from Afghanistan in 2023 as starting point, this module seeks to address what its conveners identified as a glaring gap in the academic discussion of the role of intelligence in the military. The course covered topics such as intelligence in counterinsurgency operations, UN or maritime missions, its role in tackling hybrid threats as well as in current military conflicts. Inter-active classroom discussions also reflected on the implication of how the military responds to advances in data analytics and machine learning. 

\begin{figure}[t]
\centering\includegraphics[width=\textwidth]{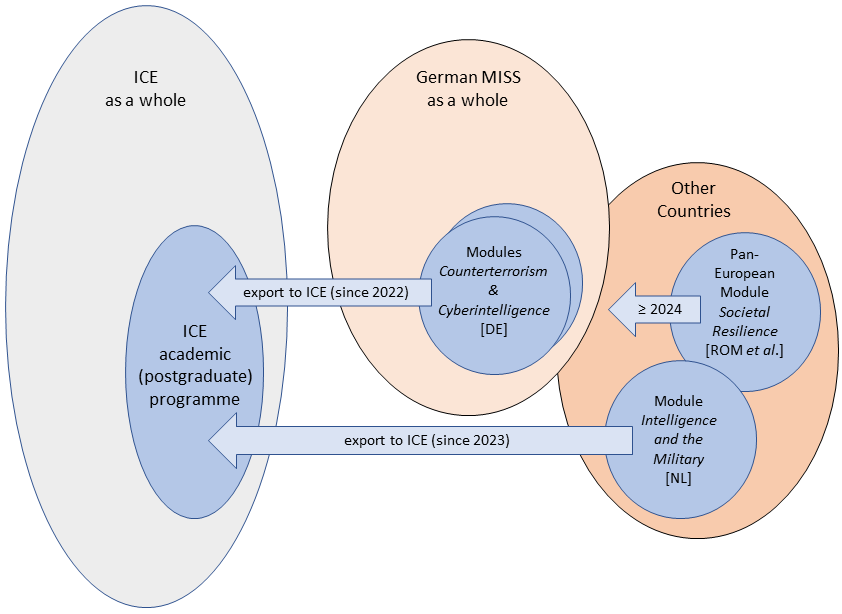}
\caption{Exports to the academic (postgraduate) programme of ICE.}\label{ICEexports}
\end{figure}

In the near future, the postgraduate programme is going to expand further with Romania leading a pan-European effort on a module {\em Societal Resilience to Hybrid Threats} that also involves contributions from Croatia, Estonia, Finland, and Germany. Figure~\ref{ICEexports} illustrates these developments.

\section*{Case Study -- The Counterterrorism Module}
The module on counterterrorism combines insights from cutting-edge research across a range of academic disciplines to provide an advanced understanding of continuities and changes in the nature of domestic and transnational terrorism, factors of radicalization, and corresponding implications for de-radicalization and counterterrorism. While the module covers all varieties of terrorist activities, it pays particular attention to current manifestations of Islamist and right-wing terrorism.

Central themes include the psychology of radicalization, determinants of the rise, decline and persistence of terrorist organizations and of factors that shape the adoption of specific terrorist tactics and strategies. Participants have the opportunity to elaborate on the most important psychological theories and models to explain radicalization, and trace individual radical biographies based on this. Moreover, discussions are aimed at assessing attempts to measure the effectiveness of terrorism and of efforts to counter its tactical and strategic impact. When considering the wider context in which terrorism occurs, participants reflect upon the changing nature in the relationship between terrorism and the media and discuss the role of active and passive state sponsors as well as of state failure in combatting radicalization and terrorism. When looking at specific manifestations of terrorism, participants discuss and reflect upon the rise of suicide and single-actor attacks, terrorist strategies like accelerationism, the possible terrorist use of weapons of mass destruction, the trans-nationalization of right- and left-wing terrorism as well as the role of women in terrorist recruitment and tactics. 

Throughout all sessions, close attention is paid to the implications for the work of and collaboration between different intelligence agencies within and across European countries. The theoretically informed and empirically enriched grasp of the current state of the art in radicalization, terrorism and counterterrorism research enables participants to evaluate the contribution of intelligence agencies to national, European, and global counterterrorism efforts.

When preparing the module for delivery as part of ICE’s postgraduate programme, the teaching team consisting of Lars Berger, Hendrik Hansen, and Michaela Pfundmair (all Federal University of Administrative Sciences, Berlin) had to pay particular attention to the question of how to ensure the pan-European appeal and relevance of content originally designed with a German intelligence and security professional in mind. The teaching team thus contemplated how a European perspective on terrorism might differ or be similar to a German perspective on terrorism and counterterrorism. Aspects that had to be considered here were different historical experiences with specific types of political extremism and terrorist violence from across the political spectrum, differences in terms of political systems and the institutional setting in which various national security agencies operate as well as the extent to which terrorism and attempts to counter it are seen as a domestic or transnational concern. When adapting the module to an international audience, Lars Berger benefited from his previous experiences in designing and implementing postgraduate programmes and modules on matters of international security at various UK universities.

In terms of the general atmosphere and setting, several observations can be made. First, participants enjoy the postgraduate atmosphere within the module. The decision to limit the number of participants to a number typical for a postgraduate seminar helped create a lively and interactive atmosphere among participants. This way, the postgraduate programme offers a nice contrast and additional dimension to other ICE events with much bigger numbers of participants. Through the additional interaction stimulated by joint presentations which are prepared during the first part of the week, participants view themselves and treat each other less as representatives of specific services and countries and more as colleagues sharing an interest in tackling a common security threat. 

Second, participants were eager to engage with academic research and the quite substantial reading list typical for postgraduate seminars. This shows that there is a real thirst among intelligence practitioners to step outside their daily demands and reflect more deeply on the security and intelligence issues they face. It this exactly this type of outside-the-box thinking stimulated by engagement with thought-provoking academic research that can assist intelligence professionals in challenging their own assumptions about the current patterns and potential future trajectories of phenomena within the security realms they monitor.

Third, participants within the module covered a broad range of backgrounds in terms of geography, experience, gender, age, and seniority. This diversity in perspectives enriched discussions enormously. When it comes to an issue such as terrorism and the question of how best to tackle it, perspectives are necessarily shaped by local manifestations of the phenomenon and historical experiences of dealing with it. Openly reflecting and critically evaluating these different perspectives marks a crucial step toward greater mutual understanding and the development of a shared European intelligence culture.

\section*{Conclusion}
The Intelligence College in Europe (ICE) provides, among others, a first response to an academic postgraduate education in intelligence and security studies on a pan-European level. It thus contributes to the formation of a European intelligence community. 

The ICE benefits from the experience of the German Master of Intelligence and Security Studies (MISS) in two ways. Firstly, the MISS shows how the various intelligence services and the military can work together in an academic educational setting. Secondly, the MISS provides two German modules (Counterterrorism / Cyberintelligence) as export products for the academic (postgraduate) ICE programme with virtually no adaptation effort. 

Using the counterterrorism module as a case study, we can vividly demonstrate the scientific depth and social relevance that can be achieved within ICE’s academic (postgraduate) programme.

Some countries have already followed this successful example and have expanded the ICE offering (or will do so soon). We hope that more players in the European intelligence community and beyond will follow suit.

\section*{About the authors}
{\bf Uwe M. Borghoff} is Vice President of the University of the Bundeswehr Munich and Director of the Campus Advanced Studies Center, the university’s center for professional education, and Director of the Center of Intelligence and Security Studies in Munich (CISS), Germany.
E-mail: uwe.borghoff@unibw.de

\noindent
{\bf Lars Berger} is professor at the Department of Intelligence at the Federal University of Administrative Sciences in Berlin, Academic Adviser to ICE as well as Germany’s academic representative within ICE.
\newline
E-mail: lars.berger@hsbund-nd.de

\noindent
{\bf François Fischer} is Director of the ICE’s Permanent Secretariat in Paris, France.
E-mail: francois.fischer@pm.gouv.fr  

%

%
%
%

\end{document}